\documentclass[a4paper,cite]{article}
\usepackage{latexsym}

\usepackage[latin1]{inputenc}

\usepackage{graphicx}
\usepackage{color,pst-plot}

\usepackage{amsmath}
\usepackage {amsfonts,amssymb,amstext}

\newcommand{\be}{\begin{equation}}
\newcommand{\ee}{\end{equation}}{
\newcommand{\bea}{\begin{eqnarray}}
\newcommand{\eea}{\end{eqnarray}}

\newcommand{\noi}{\noindent}

\DeclareMathAlphabet{\eusm}{U}{}{}{}
\SetMathAlphabet\eusm{normal}{U}{eus}{m}{n}
\SetMathAlphabet\eusm{bold}{U}{eus}{b}{n}
\DeclareMathAlphabet{\mathpzc}{OT1}{pzc}{m}{it}

\input epsf

\begin{document}

\begin{titlepage}

\begin{flushright}
CPHT-RR083.1006\\
UB-ECM-PF06/32\\
hep-ph/0611176\\

\end{flushright}

\vspace*{0.2cm}
\begin{center}
{\Large {\bf Deeply Virtual Compton Scattering on a Photon\\
and Generalized Parton Distributions in the Photon
}}\\[2 cm]

{\bf S. Friot}~$^{a}$, {\bf B. Pire}~$^b$ {\bf and  L. Szymanowski}~$^{b, c, d}$\\[1cm]

$^a$ {\it Departament d'Estructura i Constituents de la Matèria, Facultat de F\'\i sica, 08028 Barcelona, Spain}\\[0.5cm]
$^b$  {\it Centre  de Physique Th{\'e}orique, \'Ecole Polytechnique, CNRS,
   91128 Palaiseau, France}\\[0.5cm]
$^c$ {\it Université de Liège, B4000 Liège, Belgium}\\[0.5cm]
$^d$ {\it Soltan Institute for Nuclear Studies, Warsaw, Poland}
\end{center}

\vspace*{3.0cm}

\begin{abstract}
We consider  deeply virtual Compton scattering on a photon target, 
in the generalized Bjorken limit, at the Born order and in the leading logarithmic approximation.
 We  interpret 
the result as a factorized amplitude of a hard process described by  handbag diagrams and  anomalous 
generalized parton distributions in the photon.
 This anomalous part, with its characteristic $\ln (Q^2)$ dependence, is present both 
 in the DGLAP and in the ERBL regions. As a consequence, these generalized parton distributions of the photon
  obey DGLAP-ERBL evolution equations with an inhomogeneous term. 
  \end{abstract}

\end{titlepage}

\section{\normalsize Introduction}
The photon is a fascinating object for QCD studies. Among its many aspects, its parton
 distributions have been the subject of much work since the seminal paper by Witten \cite{Witten}.
 The mixing of the non local electromagnetic operator $F^{\mu\nu}(z) F^{\mu\nu}(0)$ with the 
$\bar \psi(z) \psi(0)$ quark-antiquark one, on the light-cone, 
yields quite unique features in the analysis
of the imaginary part of the forward $\gamma^* \gamma \to \gamma^* \gamma$ amplitude. 
A renormalization group analysis allows then to define
 a parton distribution in the photon which factorizes in  the structure functions measured in
 inclusive deep inelastic scattering (DIS). 
 The pointlike nature of the photon enables one to fully determine its leading  expression  in $\ln Q^2$,
 through a Born order calculation in $\alpha_{em}$ and an 
inhomogeneous QCD evolution equation which can be solved in the leading logarithmic approximation 
without assuming an additional initial condition. 

In this work we
 uncover an analogous structure in the case of the generalization of the Bjorken scaling
regime of exclusive hard reactions such  as the amplitude for  deeply virtual Compton scattering (DVCS),
$\gamma^* \gamma \to \gamma \gamma $.
 At first sight, this looks unrealistic
since QED gauge invariance demands that all diagrams contributing to the DVCS amplitude
 be considered and not all of them have a topology compatible with a partonic interpretation based on the QCD
collinear factorization of the scattering amplitude.
We show nevertheless how to define
the {\em anomalous} generalized parton distributions (GPDs) in the photon.
It can serve as a basis for
a reliable perturbative calculation of a  GPD, which is of utmost importance to test various features of 
GPDs such as sum rules, crossing properties \cite{cross} with generalized distribution 
amplitudes \cite{GDA} and positivity limits \cite{pos}.

Moreover, the parton distributions in the photon  have turned out to be of experimental importance 
in a number of accessible processes, both in $e^+ e^-$ annihilation and in photoproduction. In the same spirit
and even if DVCS on the photon seems to be more a subject for an academic study than for a phenomenological
 analysis of forthcoming data, it may  turn out that other exclusive reactions with photon GPDs 
 (e.g. $\gamma^* \gamma \to \rho \gamma$) are feasible. 
Let us stress, nevertheless, that the phenomenological use of the photon GPDs  requires 
to include  non-pointlike, hadronic contributions,  which
 effectively goes beyond the leading logarithmic approximation considered below.

\section{\normalsize The DVCS process}

\begin{figure}[h]
\centerline{\epsfxsize9.5cm\epsffile{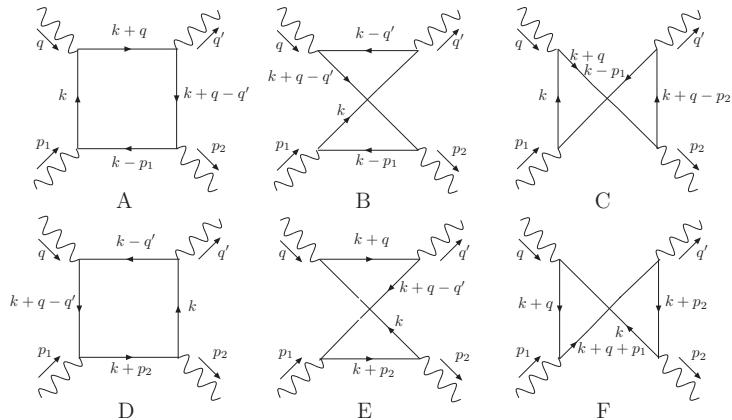}}
\caption[]{\small
The Born order diagrams for $\gamma^* \;\gamma \to \gamma  \; \gamma$ 
 }
\label{fig:1}
\end{figure}

 Deeply virtual Compton scattering  on  a photon target 
\begin{equation}
\gamma^*(q) \gamma(p_1) \to \gamma(q') \gamma(p_2)
\label{dvcs}
\end{equation}
involves, at leading order in $\alpha_{em}$, and zeroth order in  $\alpha_{S}$ the six Feynman diagrams 
of Fig.~\ref{fig:1} with quarks in the loop. 
Our aim in this section is to calculate this amplitude 
and present it in the form of an integral  over the quark momentum fraction $x$.
Since we found this calculation rather instructive, we shall describe with some details how we
 perform it on the example of the box diagram of Fig.~1A and
then exhibit  the total result coming from the sum of the six diagrams.
For simplicity, we restrict in this short paper to the case where $\Delta_T$, the transverse part of
 $\Delta\doteq p_2-p_1$, vanishes, that is zero  scattering angle (but still non-forward kinematics). We expect to  generalize the kinematics in a future work. Our conventions for the kinematics are the following:

\begin{equation}\nonumber
q=-2\xi p+n\ , ~~~~~~~~~~ q'=n\ ,
\end{equation}

\begin{equation}\nonumber
p_1=(1+\xi) p\ , ~~~~~~~~~~ p_2= p_{1} + \Delta = (1-\xi) p\ ,
\end{equation}
where $p$ and $n$ are two light-cone vectors and $2 p\cdot n =s = \frac{Q^2}{2\xi}$.

The momentum $k$ in the quark loop is chosen to be
\begin{equation}\nonumber
k^\mu \doteq(x+\xi)p^\mu +\beta n^\mu + k_T\ ,
\end{equation}
with $k_T^2=-{\bold k^2}$.

The amplitude of the process $\gamma^*\gamma\rightarrow\gamma\gamma$ with one virtual and 
three real photons  can 
be written as
\begin{equation}
A\doteq\, \epsilon_\mu\epsilon'^*_\nu{\epsilon_1}_\alpha{\epsilon^*_2}_\beta T^{\mu\nu\alpha\beta},
\end{equation}
where in our kinematics  the four
photon polarization vectors
$\epsilon(q)$, $\epsilon'^*(q')$, $\epsilon_1(p_1)$ and $\epsilon^*_2(p_2)$ are 
transverse.
The tensorial decomposition of $T^{\mu\nu\alpha\beta}$ reads \cite{BGMS75}
\begin{equation}
T^{\mu\nu\alpha\beta} (\Delta_{T}=0) = \frac{1}{4}g^{\mu\nu}_Tg^{\alpha\beta}_T W_1+
\frac{1}{8}\left(g^{\mu\alpha}_Tg^{\nu\beta}_T 
+g^{\nu\alpha}_Tg^{\mu\beta}_T -g^{\mu\nu}_Tg^{\alpha\beta}_T \right)W_2
+ \frac{1}{4}\left(g^{\mu\alpha}_Tg^{\nu\beta}_T - g^{\mu\beta}_Tg^{\alpha\nu}_T\right)W_3\, ,
\end{equation}
and it involves three scalar functions $W_i$, $i=1,2,3$.

The integration over $k$ is performed as usual within the Sudakov representation, using

$$d^4k = \frac{s}{2} dx d\beta d^2 k_{T} = \frac{\pi s}{2} dx d\beta d {\bold k^2}$$

Let us present in some details our calculations for diagram A, and afterwards only give final results for the
other diagrams. We start with the straightforward  calculation of  the imaginary  part of $W_{i}$ for this diagram.
Performing the $d\beta$ and $d{\bf k}^2$ integrations with the help of the $\delta$-functions 
defined by Cutkosky rules, we get
 for the symmetric $W_{1}$ 
 projection $g^{\mu\nu}_Tg^{\alpha\beta}_T$ :
\begin{equation}
Disc\, W_{1}^{A} =  \frac{i}{ \pi} e_{q}^4 N_{C}
\int \frac {dx}{x-\xi} \frac{(x^2+\bar x^2 -\xi^2)}{1-\xi^2}\; ,
\end{equation}
where the $x-$integration range 
is restricted by the positivity conditions attached to the $\delta$-functions,
to 
\begin{equation}
\xi +\frac{m^2}{s} < x <  1-\frac{m^2}{s}\;.
\end{equation}
This leads for this diagram to
\begin{equation}
Disc\, W_{1}^{A}  =  2i{\cal I}m\, { W}_1^A = - \frac{i}{ \pi} e_{q}^4 N_{C}
  \frac{(1 -\xi)^2}{1-\xi^2} \log \frac{m^2}{Q^2} \; ,
\end{equation}
where we restricted ourselves to the logarithmic terms and traded $s$ for  $Q^2$ in the argument of the logarithm.

The corresponding result for the  discontinuity of $W_{3}$ attached to the $\mu \leftrightarrow \nu$ 
antisymmetric tensorial structure  is
\begin{equation}
Disc\, W_{3}^{A} =  2i{\cal I}m\, W_3^A = \frac{i}{ \pi} e_{q}^4 N_{C}
  \frac{(1 -\xi)^2}{1-\xi^2} \log \frac{m^2}{Q^2} \;,
\end{equation}
and the third discontinuity  $Disc\, W_{2}^{A}$     vanishes.

The calculation of the real part ${\cal R}e\, W_{1}^{A}$ is more involved. We get for ${\cal R}e\, W_{1}^{A}$
\begin{equation}
 {\cal R}e\; i \frac{s e_{q}^4 N_{C}}{32 \pi^3} \int
  \frac{dx\; d\beta\; d{\bold k^2}\;\;\;Tr A_{1}}{[(k+q)^2 -m^2+i\eta] [(k-p_{1})^2-m^2+i\eta](k^2-m^2+i\eta)[(k+\Delta)^2-m^2+i\eta]}\,,
\label{ReW}
\end{equation}
where $Tr A_1 = Tr (\gamma_{T}^\nu (\hat k + \hat q + m) \gamma_{T}^\nu(\hat k+m)
\gamma_{T}^\alpha (\hat k -\hat p_{1}+m)\gamma_{T}^\alpha (\hat k +\hat q -\hat n +m)\;
 $.

We first integrate in $\beta$ using the Cauchy theorem. The propagators induce poles in the complex $\beta$-plane
with values
\begin{eqnarray}
\beta_{1} &=&   \frac {{\bold k^2} + m^2 -i\eta}{s(x+\xi)} ~~~~~~~~~~\beta_{2}= 
-1 + \frac {{\bold k^2} + m^2 -i\eta}{s(x-\xi)}\nonumber\\
\beta_{3}&=&   \frac {{\bold k^2} + m^2 -i\eta}{s(x-\xi)} ~~~~~~~~~~\beta_{4}= 
\frac {{\bold k^2} + m^2 -i\eta}{s(x-1)} \;.
\end{eqnarray}
Since the four poles are all below the real axis for $x>1$ and all above the real 
axis for $x<-\xi$, the only region 
where the amplitude may not vanish is   $1 > x > -\xi$. 
In this way we identify two different regions  : 
\begin{itemize}
\item  the  DGLAP  region where  $\xi<x<1$, for which one may close the contour 
in the upper half plane and get the contribution of the pole $\beta_{4}$\,,
\item   the  ERBL region where  $-\xi<x<\xi$, for which one may close the contour 
in the lower half plane and get the contribution of the pole $\beta_{1}$\,.
\end{itemize}
A number of technical simplifications are now helpful. 
The integral over ${\bold k^2}$ in Eq.~(\ref{ReW}) is UV divergent. On the other hand, it is 
a well-known classical result of QED  that the sum of integrals 
corresponding to the six diagrams of Fig. 1 is finite, so we will separate UV divergent terms of each diagram in 
an algebraic way.  
One easily verifies that, at the leading logarithmic approximation 
we are interested in, the trace may be simplified by taking the limit $m^2\to 0$.
  The integral in Eq.~(\ref{ReW}) may then be written as
\begin{equation}
I^{A} = I^{A}_{4}+ I^{A}_{1}= 2i\pi\int_{\xi}^1 dx\, \int\, d{\bold k^2} \frac{Tr A _1(\beta = \beta_{4})}{DA_{\beta 4}} -2i\pi
\int_{-\xi}^\xi dx\, \int\, d{\bold k^2} \frac{Tr A_1 (\beta = \beta_{1})}{DA_{\beta 1}}\;,
\end{equation}
with the pole values of the propagators :
\begin{eqnarray}
DA_{\beta 4} &=&  \frac{s ({\bold k^2} + m^2)^2 (1-\xi^2) ((1-\xi)({\bold k^2} + m^2) + s(1-x)(\xi-x))}{(1-x)^2} \nonumber\\
DA_{\beta 1} &=&  \frac{-2s ({\bold k^2} + m^2)^2 \xi(1+\xi) (2\xi({\bold k^2} + m^2) + s(\xi^2-x^2))}{(\xi+x)^2} \; . \nonumber
\end{eqnarray}
 The traces are simple polynomials in ${\bold k^2}$, which may be written as ($i = 1, 4$)
 $Tr A_{1} (\beta = \beta_{i})\sim \alpha_{i}{\bold k^4} + \gamma_{i}Q^2 {\bold k^2}+ \delta_{i} Q^4$.
Power counting in ${\bold k^2}$ shows that the $\alpha_{i}{\bold k^4}$ term in these integrals is ultraviolet divergent.  
Our aim is to recover the UV finiteness of the sum of diagrams 
before performing the integration over $x$. For this, we simply separate
the ultraviolet divergent part of the integrand,  $\frac {\alpha_{i}{\bold k^4}}{DA_{\beta i}}$. 
This yields for the diagram A and the DGLAP region $\xi<x<1$ (at the leading logarithm level)
\begin{eqnarray}
I^{A}_{4} = 2i\pi\int_{\xi}^1 dx\, \int \,  d{\bf k}^2 \frac{Tr A^r_1 (\beta = \beta_{4})}{DA_{\beta 4}} 
+ 2i\pi \int_{\xi}^1 dx  \int d{\bold k^2}\frac{\alpha_{4}{\bold k^4}}{DA_{\beta 4}}\;,
\end{eqnarray}
with the regularized numerator $Tr A^r_1 (\beta = \beta_{4})= \gamma_{4}Q^2 {\bold k^2} + \delta_{4} Q^4$.
Simple algebra leads then to:
\begin{equation}
I^{A}_{4} = \int_{\xi}^1 dx  \frac{2(x^2+(1-x)^2-\xi^2)}{\pi\,s\,(1-\xi^2)(\xi -x)} \log \frac{m^2}{Q^2}  + 
 2i\pi \int_{\xi}^1 dx  \int d{\bold k^2}\frac{\alpha_{4}{\bold k^4}}{DA_{\beta 4}} \; .
\end{equation}
The corresponding result for the ERBL region is
\begin{equation}
I^{A}_{1} = \int_{-\xi}^{\xi} dx  \frac{(x +\xi)(\xi -2x +1)}{\pi\,s\,\xi(1+\xi)(x-\xi)} \log \frac{m^2}{Q^2}  + 
 2i\pi \int_{-\xi}^{\xi} dx  \int d{\bold k^2}\frac{\alpha_{1}{\bold k^4}}{DA_{\beta 1}} \; .
\end{equation}
The same procedure is applied to the five other diagrams of Fig.~1. Diagrams B and C 
contribute to the same regions  $\xi<x<1$ 
and  $-\xi<x<\xi$ as  diagram A.
Diagrams D, E, F contribute to the {\em "antiquark"} regions  $-1<x<-\xi$ and  $-\xi<x<\xi$. The  results 
for diagrams B, D and E are similar to the one for diagram A. 
It is worth emphasizing that  diagrams C and F 
in our leading logarithmic  approximation contribute only  to the  ultraviolet divergent 
contribution and {\em not} to any 
$\log  \frac{m^2}{Q^2}$ terms. This fact will lead to a 
handbag dominance interpretation of the leading logarithmic 
result, as discussed later on.

At this point, we observe that  the ultraviolet divergent contributions cancel out, 
{\em before} integration in the $x$ variable, and moreover separately 
in the two sets of diagrams  (A, B, C) and (D, E, F).

The final result of our calculation of the real part of the DVCS amplitude reads
\begin{eqnarray}
{\cal R}e\, W_{1}  &=& \frac{e_q^4\,N_C}{2\,\pi^2}\int_{-1}^{1} \,dx
\frac{2\,x}{(x-\xi)(x+\xi)}\left[\theta(x-\xi)  
\frac {x^2 + (1-x)^2-\xi^2}{1-\xi^2}\right.\;\nonumber\\
&+& \left. \theta(\xi-x) \theta(\xi+x) \frac{x(1-\xi)}{\xi(1+\xi)} 
- \theta(-x-\xi)  \frac {x^2 + (1+x)^2-\xi^2}{1-\xi^2}\;\right] \ln\frac{m^2}{Q^2}\;.
\label{RW1}
\end{eqnarray}
The real parts of  $W_{2}$ and $W_{3}$  are calculated in the same way. We get
\begin{equation}
{\cal R}e\, W_{2} =0
\label{RW2}
\end{equation}
and
\begin{eqnarray}
{\cal R}e\, W_{3}  &=& \frac{e_q^4\,N_C}{2\,\pi^2}\int_{-1}^1 \,dx 
\frac{2\xi}{(x-\xi)(x+\xi)}
\left[\theta(x-\xi)   \frac {x^2 - (1-x)^2-\xi^2}{1-\xi^2}\; \right.\nonumber\\
 &-& \left. \theta(\xi-x) \theta(\xi+x) \frac{1-\xi}{1+\xi} \;
+ \theta(-x-\xi)  \frac {x^2 - (1+x)^2-\xi^2}{1-\xi^2}\;\right] \ln\frac{m^2}{Q^2}\;.
\label{RW3}
\end{eqnarray}
The real and imaginary parts of the amplitudes may be put together in the simple form :
\begin{eqnarray}
 W_{1}  &=& \frac{e_q^4\,N_C}{2\,\pi^2}\int_{-1}^1\,dx
\frac{2\,x}{(x-\xi+i\eta)(x+\xi-i\eta)}
 \left[\theta(x-\xi)  \frac {x^2 + (1-x)^2-\xi^2}{1-\xi^2}\; \right.\nonumber\\
&+& \left.\theta(\xi-x) \theta(\xi+x) \frac{x(1-\xi)}{\xi(1+\xi)} 
- \theta(-x-\xi)  \frac {x^2 + (1+x)^2-\xi^2}{1-\xi^2}\;\right] \ln\frac{m^2}{Q^2}
\label{W1}
\end{eqnarray}
and
\begin{eqnarray}
 W_{3}  &=& \frac{e_q^4\,N_C}{2\,\pi^2}\int_{-1}^1\,dx
\frac{2\xi}{(x-\xi+i\eta)(x+\xi-i\eta)}
 \left[\theta(x-\xi)   \frac {x^2 - (1-x)^2-\xi^2}{1-\xi^2}\; \right. \nonumber\\
&-& \left. \theta(\xi-x) \theta(\xi+x) \frac{1-\xi}{1+\xi} \;
+ \theta(-x-\xi)  \frac {x^2 - (1+x)^2-\xi^2}{1-\xi^2}\;\right] \ln\frac{m^2}{Q^2}\;.
\label{W3}
\end{eqnarray}
We now want to interpret this result from the point of view of QCD factorization based on the operator product expansion, 
yet still in the zeroth order of the QCD coupling constant and in the leading logarithmic approximation.
 For this, we write for any function ${\cal F}(x,\xi)$ the obvious identity :
 \begin{equation}
{\cal F}(x,\xi) \; \log \frac{m^2}{Q^2} = {\cal F}(x,\xi) \;\log \frac{m^2}{M_{F}^2} + {\cal F}(x,\xi) \;\log \frac{M_{F}^2}{Q^2}\;,
\label{stupid}
\end{equation}
where $M_{F}$ is an arbitrary factorization scale. We will show below that the first term with $\log \frac{m^2}{M_{F}^2}$
may be identified with the quark content of the photon, 
whereas the second term with $\log \frac{M_{F}^2}{Q^2}$
corresponds to the so-called photon content of the photon, coming from the  matrix element of
the two photon correlator $A_\mu(-\frac{z}{2}) A_\nu (\frac{z}{2})$ which contributes at the same order in
$\alpha_{em}$ as the quark correlator to the scattering amplitude. 
Choosing $M_{F}^2= Q^2$ will allow to express the DVCS amplitude only 
in terms of the quark content of the photon.

\section{\normalsize  QCD factorization of the DVCS amplitude on the photon}

To understand the results of Eqs.~(\ref{W1}-\ref{W3}) within the QCD factorization, we  first consider 
two quark non local correlators on the light cone and their matrix elements between real photon states :

\begin{equation}
\label{Fqa}
F^q = \int \frac{dz}{2\pi} e^{ixz}\langle \gamma(p')| \bar q(-\frac{z}{2}N)
 \gamma.N q(\frac{z}{2}N)|\gamma(p) \rangle 
\end{equation}
and
\begin{equation}
\tilde F^q = \int \frac{dz}{2\pi} e^{ixz}\langle \gamma(p')| \bar q(-\frac{z}{2}N)
 \gamma.N \gamma^5 q(\frac{z}{2}N)|\gamma(p) \rangle 
\end{equation}
where we note $N= n/n.p$ and where we neglected, for simplicity of notation, both 
the electromagnetic and the gluonic Wilson lines.

There also exists the photon 
 correlator $ F ^{N \mu}(-\frac{z}{2}N) F^{\nu N}(\frac{z}{2}N)$ (where $F^{N \mu} = N_\nu F^{\nu \mu}$), which mix  with the quark operators \cite{Witten}, but 
contrarily to the quark correlator matrix element, the photonic one begins at order $\alpha_{em}^0$,
as seen  for instance in the symmetric case :
\begin{equation}
\label{FF}
\int \frac{dz}{2\pi} e^{ixz}\langle \gamma(p_{2})| F ^{N \mu}(-\frac{z}{2}N)
F^{\nu N}(\frac{z}{2}N) g_{T\mu\nu}|\gamma(p_{1}) \rangle = -  g_T^{\mu\nu}
\epsilon_\mu (p_{1})\epsilon^*_\nu(p_{2}) (1-\xi^2) [\delta(1+x) + \delta(1-x)] \; .
\end{equation}

The quark correlator matrix elements, calculated in the lowest order of $\alpha_{em}$ and $\alpha_{S}$, 
suffer from ultraviolet divergences, which we regulate 
through the usual dimensional regularization procedure, with  $d= 4+2\epsilon$. 
We obtain (with  $\frac{1}{\hat\epsilon} = \frac{1}{\epsilon} +\gamma_{E}-\log 4\pi$)
\begin{equation}
F^q= \frac{N_C\,e_{q}^2}{4\pi^2} g_T^{\mu\nu}\epsilon_\mu (p_{1})\epsilon^*_\nu(p_{2}) 
 \left[\frac{1}{\hat\epsilon} + \log{m^2}\right] F(x,\xi)\,,
\label{Fgam}
\end{equation}
with 
\begin{equation}
 F(x,\xi) = \frac{x^2+(1-x)^2-\xi^2}{1-\xi^2}\theta(1>x>\xi) 
 -\frac{x^2+(1+x)^2-\xi^2}{1-\xi^2}\theta(-\xi>x>-1) 
 +\frac{x(1-\xi)}{\xi+\xi^2}\theta(\xi>x>-\xi)
\end{equation}
for the $\mu\leftrightarrow \nu$ symmetric (polarization averaged) part. The corresponding results for 
 the antisymmetric (polarized) part read
\begin{equation}
\tilde F^q= \frac{N_C\,e_{q}^2}{4\pi^2}(-i \epsilon^{\mu\nu p N})\epsilon_\mu (p_{1})\epsilon^*_\nu(p_{2}) 
 \left[\frac{1}{\hat\epsilon}  + \log m^2\right] \tilde F(x,\xi)
\label{Ftgam}
\end{equation}
with
\begin{equation}
 \tilde F(x,\xi) = \frac{x^2-(1-x)^2-\xi^2}{1-\xi^2}\theta(1>x>\xi) 
 +\frac{x^2-(1+x)^2-\xi^2}{1-\xi^2}\theta(-\xi>x>-1) 
 -\frac{1-\xi}{1+\xi}\theta(\xi>x>-\xi)\; .
\end{equation}
 Let us stress again that here we concentrate only on the leading logarithmic behaviour and thus
 focus on the divergent parts and their associated logarithmic functions. The ultraviolet divergent parts
 are removed through the renormalization procedure  involving both quark and photon correlators 
   (see for example \cite{Hill}). 
  The $\frac{1}{\hat \epsilon}$ terms in 
Eqs.~(\ref{Fgam}, \ref{Ftgam}) define the 
non-diagonal element $Z_{\bar q  q\;F
F}$ of the multiplicative matrix of  renormalization constants $Z$. 
The $\frac{1}{\hat \epsilon}$ terms are
then subtracted by the renormalization of quark operators.
The renormalization procedure introduces a renormalization scale which we here identify 
with a factorization scale  $M_{F }$ in the factorized form of the amplitude 
(see Eq.~(\ref{fac}) below). Imposing the renormalization condition that the renormalized 
quark correlator matrix element vanishes
when the factorization scale $M_{F } = m$, we get from Eq.~(\ref{Fgam}) for the renormalized matrix element
(\ref{Fqa})
\begin{equation}
F^q_{R} 
 = \frac{N_C\,e_{q}^2}{4\pi^2} g_T^{\mu\nu}\epsilon_\mu (p_{1})\epsilon_\nu(p_{2}) 
  \log{\frac{m^2}{M_{F}^2}} F(x,\xi) \; ,
\label{FRgam}
\end{equation}
and a similar result for the antisymmetric case of Eq.~(\ref{Ftgam}):
\begin{equation}
\tilde F^q_{R}
 = \frac{N_C\,e_{q}^2}{4\pi^2}(-i  \epsilon^{\mu\nu p N})\epsilon_\mu (p_{1})\epsilon_\nu(p_{2}) 
  \log{\frac{m^2}{M_{F}^2}} \tilde F(x,\xi) \; .
\label{FRtgam}
\end{equation}
Eqs.~(\ref{FRgam}, \ref{FRtgam}) permit us to define the generalized quark distributions in the photon, $H_i^q(x,\xi,0)$, as
\begin{equation}
\label{Hquark}
F^q_{R}= - g_T^{\mu\nu}\epsilon_\mu (p_{1})\epsilon^*_\nu(p_{2}) H_{1}^q (x,\xi,0)\;,\;\;\;\;
\tilde F^q_{R}=  i\epsilon^{\mu\nu p N}\epsilon_\mu (p_{1})\epsilon^*_\nu(p_{2}) H_{3}^q (x,\xi,0)\;.
\end{equation}
and to 
write the quark contribution to the DVCS amplitude as a convolution of coefficient functions and 
distributions $ H_{i}^q$  
\begin{equation}
 W^q_{1}= \int\limits_{-1}^1 dx C_V^q(x)  H_{1}^q (x,\xi,0)\;\,,~~~~~~~~~~W^q_{3}= \int\limits_{-1}^1 dx 
C_A^q(x)  H_{3}^q (x,\xi,0)\; ,
\label{fac}
\end{equation}
where the  Born order coefficient functions $C_{V/A}^q$ attached to the quark-antiquark symmetric and 
antisymmetric correlators are the usual hard process amplitudes :
 \begin{equation}
 C_{V/A}^q = - 2e_q^2\left(\frac{1}{x-\xi+i\eta} \pm \frac{1}{x+\xi-i\eta}\right)\; .
\end{equation}
We recover in that way the $\ln \frac{m^2}{M_{F}^2}$ term in the right hand side of Eq.~(\ref{stupid}).

The photon operator contribution to the DVCS amplitude at the order $\alpha_{em}^2$ considered here,
involves  a new coefficient function calculated at the factorization scale $M_{F}$, which
 plays the role of the infrared cutoff, convoluted with the
photon correlator (\ref{FF}) or with its antisymmetric counterpart. 
The results of these convolutions effectively coincide
with the amplitudes calculated in Section 2 with the quark mass replaced by the factorization scale 
$m \to M_{F}$, and leads to the second term in the right hand side of Eq.~(\ref{stupid}).
The triviality of Eq.~(\ref{stupid}) in fact hides the more general independence of the scattering amplitude
 on the choice of the scale  $M_{F}$ which is controlled by the renormalization group equation.

We still have the freedom to fix the factorization scale $ M_{F}^2$ in any convenient way.
Choosing $ M_{F}^2 = Q^2$ kills the logarithmic terms coming from the photon correlator, so that the 
DVCS amplitude is written (at least, in the leading logarithmic approximation) solely in terms of the quark correlator,
recovering a partonic interpretation of the process, see e.g. \cite{deWitt}.

\section{\normalsize The GPDs of the photon and their QCD evolution equations}

We have thus demonstrated that it is legitimate to define the  Born order generalized quark distributions  
in the photon at zero $\Delta_{T}$ as
\begin{eqnarray}
\label{H1}
H_{1}^q (x, \xi, 0) &=&\frac{N_C\,e_{q}^2}{4\pi^2} \left[\theta(x-\xi)  \frac {x^2 + (1-x)^2-\xi^2}{1-\xi^2}\;\right. \nonumber\\
 &+& \left.\theta(\xi-x) \theta(\xi+x) \frac{x(1-\xi)}{\xi(1+\xi)} \;
- \theta(-x-\xi)  \frac{x^2+(1+x)^2-\xi^2}{1-\xi^2}\right]\; \ln\frac{Q^2}{m^2}
\end{eqnarray}
\begin{eqnarray}
\label{H3}
H_{3}^q (x, \xi, 0) &=& \frac{N_C\,e_{q}^2}{4\pi^2} \left[\theta(x-\xi)   \frac {x^2 - (1-x)^2-\xi^2}{1-\xi^2}\;\right. \nonumber\\
 &-& \left.\theta(\xi-x) \theta(\xi+x) \frac{1-\xi}{1+\xi} \;
+ \theta(-x-\xi)  \frac {x^2 - (1+x)^2-\xi^2}{1-\xi^2}\right]\; \ln\frac{Q^2}{m^2}\;.
\end{eqnarray}
Since we focussed on the logarithmic factors, we only 
obtained the {\em anomalous} part of these GPDs. Their $x-$ and $\xi-$dependence  are shown on Figs. 2 and 3.
The $\xi \to 0$ limit of Eqs.~(\ref{H1}, \ref{H3}) leads to the usual anomalous 
Born order quark content of the photon. The $H_i^q(x,\xi,0)$ GPDs are continuous functions of $x$ at 
the points $x=\pm \xi$, but their derivatives are not.

On can verify that these GPDs obey the usual polynomiality conditions
\begin{eqnarray}
\int\, dx\,x^{2p+1} \, H_{1}^q (x, \xi, 0) &\sim & 
\frac{1}{p+1}-\frac{1}{(2p+3)(p+2)}\left[\frac{3p+4}{p+1}\xi^{2p+2}+2\sum_{m=0}^{p}\xi^{2m}\right]     
\; \nonumber\\
\int\,dx\, x^{2p}\, H_{3}^q (x, \xi, 0) &\sim & 
\frac{2}{2p+1}-\frac{2}{(2p+1)(p+1)}\sum_{m=0}^{p}\xi^{2m}     \;,
\end{eqnarray}
the other moments vanish.

\begin{figure}[h]
\centerline{\epsfxsize9.5cm\epsffile{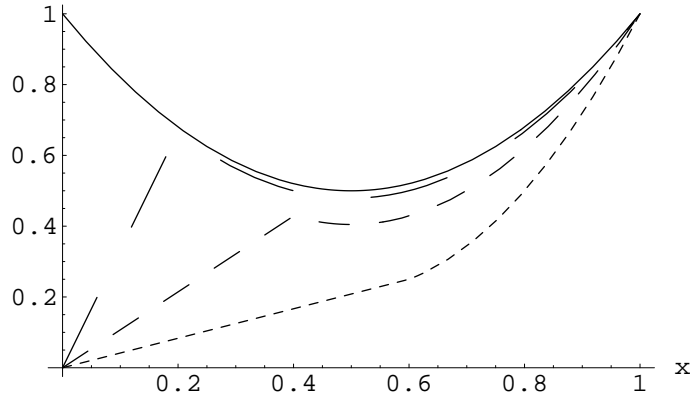}}
\caption[]{\small
The unpolarized anomalous photon GPD $H_1^q/(N_C\,e_q^2/(4\pi^2)\,\ln \frac{Q^2}{m^2})$ at Born order and $\Delta_T=0$
for $\xi=0$ (full line), $0.1$ (long dashed line), $0.3$ (middle 
dashed line), $0.5$ (small dashed line). }
\label{fig:2}
\end{figure}
\begin{figure}[h]
\centerline{\epsfxsize9.5cm\epsffile{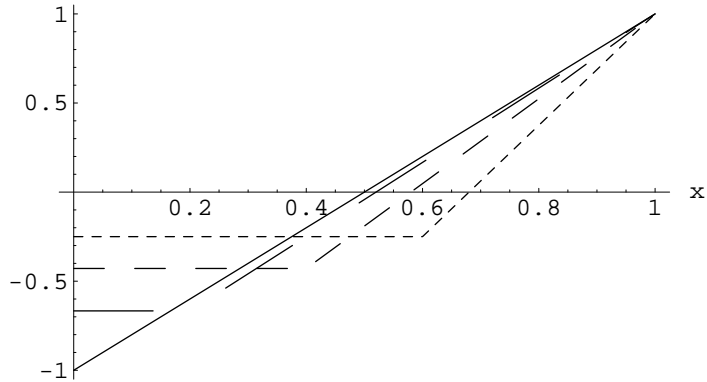}}
\caption[]{\small
The  polarized anomalous photon GPD $H_3^q/(N_C\,e_q^2/(4\pi^2)\,\ln \frac{Q^2}{m^2})$ at Born order and $\Delta_T=0$ 
for $\xi=0$ (full line), 
$0.1$ (long dashed line), $0.3$ (middle dashed line) 
$0.5$ (small dashed line). }
\label{fig:3}
\end{figure}
Switching on QCD, the photon GPDs evolve from this QED-like parton model result
 through  modified DGLAP-ERBL  evolution equations
  \begin{equation}  
\mu^2\,\frac{d}{d\,\mu^2} H_{i}^q(x,\xi, 0)= f_i^q(x,\xi)+\int\limits_{-1}^1\,dx'
\frac{1}{\xi}\,V_{NS}\left( \frac{x}{\xi},\frac{x'}{\xi} \right) 
H_{i}(x',\xi,0) \;,
\label{eveq}
\end{equation} 
where $f_i^q(x,\xi)$ are defined by the r.h.s. of  
 Eqs.~(\ref{H1}, \ref{H3}), as the corresponding functions which 
multiply $\ln Q^2/m^2$. The QCD kernel $V_{NS}$ is well-known \cite{GPD} to the leading order accuracy
\begin{eqnarray}
&&V_{NS}(x,x') = 
\frac{\alpha_s}{4\,\pi}C_F \left[\rho(x,x')\left[ 
\frac{1+x}{1+x'}\left( 1+\frac{2}{x'-x}  \right) \right]  + [x\to -x, x' 
\to -x']  \right]_+ \nonumber \\
&& \rho(x,x')=\theta(x'\geq x \geq -1) - \theta(x' \leq x \leq -1)\;,
\label{kernel}
\end{eqnarray} 
with $C_F=(N_C^2-1)/(2N_C)$. 
As in the forward case, the presence of the inhomogeneous term allows to fully evaluate the GPDs. 
We will address this point in a future work.

\section{\normalsize  Conclusions}

We derived the leading amplitude of the DVCS (polarization averaged or polarized) process on
a photon target. We have shown that the amplitude coefficients $W_i^q$ factorize in the forms shown in 
Eq.~(\ref{fac}),
irrespectively of the fact that the handbag diagram interpretation appears only 
{\em after} cancellation of UV divergencies in the scattering amplitude.
We have shown that the objects $H_i^q(x,\xi,t)$ are matrix elements of non-local quark operators 
on the light cone, and that they have an anomalous component which is proportional to $\log(Q^2/m^2)$.
They thus have all the properties attached to generalized parton distributions, and 
they obey {\em non-homogeneous} DGLAP-ERBL evolution equations. 

The extension of the results of this work beyond the leading logarithmic approximation requires 
taking into account the non-pointlike, hadronic contribution of the photon wave function.
In the diagonal case of the photon structure function, 
it is done in a model dependent way, 
by invoking the vector meson dominance hypothesis (see second Ref. \cite{Witten}).  
This approach encounters, however, well known difficulties related to the impossibility to separate
in a precise way those non-pointlike contributions from the 
 next to leading order corrections to the processes with  the pointlike coupling of the photon \cite{a+g}.
In the non-diagonal case considered here, we may apply a similar strategy and define non-pointlike 
photon GPDs by relating them to the vector meson GPDs or to the $\gamma \to$ vector meson 
transition distribution amplitude \cite{tda}.

\section*{Acknowledgments}

\noi
We are grateful to Igor Anikin, 
Markus Diehl, Thorsten Feldmann, Georges Grunberg, Dmitri Ivanov, Michael Klasen, Jean-Philippe Lansberg and Samuel Wallon for useful 
discussions and correspondance. 
This work has been completed during the program ``From RHIC to LHC: Achievements and 
Opportunities (INT-06-3)'' at Institute for Nuclear Theory in Seattle. L.S. thanks INT for hospitality. 
S.F. would like to thank \'Ecole Polytechnique for 
giving him the opportunity to work in the CPhT during the year 2005-2006.
This work is partly supported by the French-Polish scientific agreement Polonium, 
the Polish Grant 1 P03B 028 28,  the ECO-NET program, contract 
12584QK, the Joint Research Activity "Generalised Parton Distributions" of the european I3 program
Hadronic Physics, contract RII3-CT-2004-506078 and the FNRS (Belgium).

\end{document}